\documentclass{jpp}
\usepackage{graphicx}

\usepackage[utf8]{inputenc}
\usepackage[T1]{fontenc}
\usepackage{amsmath}
\usepackage{amssymb}
\usepackage{physics}
\usepackage{siunitx}
\AtBeginDocument{\RenewCommandCopy\qty\SI}
\usepackage{dcolumn}
\usepackage{bm}
\usepackage{xcolor}
\usepackage[colorinlistoftodos]{todonotes}
\usepackage{soul}
\usepackage{xfrac} 

\usepackage[pdfusetitle]{hyperref} 
\hypersetup{colorlinks, 
            citecolor=blue,
            linkcolor=blue,
    		urlcolor=blue}
      
\usepackage{float} 	
\usepackage{microtype} 
\usepackage{subcaption}	
\usepackage{booktabs}
\usepackage[noabbrev,nameinlink]{cleveref} 

\DeclareUnicodeCharacter{03B2}{{$\beta$}}
\DeclareUnicodeCharacter{2207}{{$\nabla$}}

\let\vec\bm 

\shorttitle{Pellet rocket effect with plasmoid drifts}
\shortauthor{N.~J.~Guth et al.}

\title{The effect of plasmoid drifts on the pellet rocket effect in magnetic confinement fusion plasmas}

\author{N.~J.~Guth\aff{1,2}
  \corresp{\email{nico.guth@ipp.mpg.de}}, O.~Vallhagen\aff{1}, P.~Helander\aff{2},  A.~Tresnjic\aff{1}, S.~L.~Newton\aff{5}, T.~F\"ul\"op\aff{1,4,5},   I.~Pusztai\aff{1}}

\affiliation{\aff{1}Department of Physics, Chalmers University of Technology,  G\"{o}teborg, SE-41296, Sweden
\aff{2} Max Planck Institute for Plasma Physics, 17491 Greifswald, Germany
\aff{3} United Kingdom Atomic Energy Authority, Culham Campus, Abingdon, Oxon OX14 3DB, UK
\aff{4} Rudolf Peierls Centre for Theoretical Physics, University of Oxford, Oxford, OX1 3PU, UK
\aff{5} Merton College, Oxford, OX1 4JD, UK
}

\begin{document}

\maketitle

\begin{abstract}
We detail here a semi-analytical model for the \emph{pellet rocket effect}, which describes the acceleration of pellets in a fusion plasma due to asymmetries in the heat flux reaching the pellet surface and the corresponding ablation rate. 
This effect was shown in experiments to significantly modify the pellet trajectory, and projections for reactor scale devices indicate that it may severely limit the effectiveness of pellet injection methods. We account for asymmetries stemming both from plasma parameter gradients and an asymmetric plasmoid shielding caused by the drift of the ionized pellet cloud. For high temperature, reactor relevant scenarios, we find a wide range of initial pellet sizes and speeds where the rocket effect severely limits the penetration depth of the pellet. In these cases, the plasma parameter profile variations dominate the rocket effect. We find that for small and fast pellets, where the rocket effect is less pronounced, plasmoid shielding induced asymmetries dominate.   

\end{abstract}


\section{Introduction}
\label{sec:introduction}

Injection of fuel or impurities in the form of small cryogenic pellets is an essential tool to sustain and control magnetic confinement fusion plasmas. The deposition of cold, dense material in this way may be used for fueling, tailoring plasma parameter profiles, controlling edge localized modes and mitigating disruptions \citep{pegourie_review_2007}, as well as for diagnostic purposes \citep{kuteev_impurity_1994}. In the form of shattered pellet injection (SPI) \citep{commaux_demonstration_2010}, where a larger pellet is broken into a plume of shards before entering the plasma, it is foreseen as the main disruption mitigation system in ITER \citep{LehnenITER}. Especially for disruption mitigation, besides the requirement of a rapid delivery of the injected material, it is important that the material reaches deep into the plasma.   

A pellet traversing the magnetically confined plasma is continuously heated by incident hot electrons, leading to an ablation of the pellet surface. The ablated material forms a cold, dense neutral cloud around the pellet, which then absorbs most of the incoming heat. The shielding is a self-regulated process, balancing the rate of ablation and the opacity of the ablation cloud.
Close to the pellet the ablated material forms a neutral gas, while further away it ionizes, forming a so-called plasmoid.

The plasmoid expands along the magnetic field lines, and drifts due to the inhomogeneous magnetic field \citep{lang_high-efficiency_1997,parks_radial_2000} -- towards the low-field side (LFS) in a tokamak.
Eventually, the pellet material homogenizes over the flux surfaces and equilibrates with the background plasma.
An accurate prediction of the resulting material deposition profile is essential, as it provides the basis for the design of pellet injection scenarios in future fusion reactors.
Most previous modelling efforts have focused on the rate of ablation and the homogenization processes \citep{pegourie_review_2007}, while treating the pellet motion as uniform and linear.

It was apparent already from the first pellet injection experiments that pellets can be deflected in the toroidal direction \citep{jorgensen_ablation_1975,foster_solid_1977,combs_pellet_1993}.
It was soon established that this is caused by an asymmetric heating of the pellet, which intensifies the ablation over some region of the pellet surface and propels it in the opposite direction \citep{jones_self-propelling_1978,andersen_injection_1985} -- a phenomenon appropriately termed the \emph{pellet rocket effect}. 

In tokamaks, the ultimate cause of the heating asymmetry that causes a \emph{toroidal} deflection is the plasma current \citep{andersen_injection_1985,kuteev_hydrogen_1995,waller_investigation_2003}. In addition, pellet acceleration towards the LFS has been observed in tokamaks such as ASDEX \citep{wurden_pellet_1990}, ASDEX Upgrade \citep{muller_high-_1999,kocsis_fast_2004}, HL-1M \citep{liu_overview_2002} and JET \citep{jachmich_shattered_2022, kong_interpretative_2024}. This radial pellet rocket effect has also been observed in stellarators such as TJ-II \citep{medina-roque_studies_2021}, LHD \citep{mishra_observation_2011} and Wendelstein-7X \citep{baldzuhn_pellet_2019}.
In some stellarator studies, toroidal and poloidal rocket acceleration has been connected to energetic ions introduced by neutral beam injection \citep{morita_observation_2002,matsuyama_over-ablation_2012,panadero_experimental_2018}.

While a drag force between the drifting plasmoid and the neutral gas has been considered as a possible factor in the radial pellet acceleration \citep{polevoi_simplified_2001}, the dominant mechanism is likely to be the rocket effect.
The required heating asymmetry arises due to the radial variation of plasma parameters, as well as asymmetric heat flux attenuation by the plasmoid, caused by the plasmoid drift.
The latter has been modelled semi-analytically by \citet{senichenkov_pellet_2007}, connecting the induced asymmetry in the ablation rate with the rocket force, while neglecting the pressure asymmetry at the pellet surface.
A semi-empirical model which depends on this pressure difference was developed by \citet{szepesi_radial_2007}, where the pressure asymmetry is treated as a free parameter.

A recent 3D Lagrangian particle code simulation study by \citet{samulyak_simulation_2023,samulyak_lagrangian_2021}, which self-consistently computes many aspects of the pellet deposition process, including the pressure asymmetry, suggests that the radial pellet rocket effect yields acceleration values that can significantly affect pellet trajectories in ITER.
Motivated by the potential importance of the rocket effect, we have developed a semi-analytical model to describe it, suitable for implementation into reduced numerical models. 
We treat the asymmetries as perturbations around the spherically symmetric solution of the widely used neutral gas shielding (NGS) model developed by \citet{parks_effect_1978}.
The pressure asymmetry is thus self-consistently calculated, in response to asymmetric heating boundary conditions.

The model is valid for any arbitrary source of electron heating asymmetry onto hydrogenic pellets, and was first presented in  
\citep{GuthPRL}, where the rocket effect induced by the gradients in the background plasma parameters was studied.
The purpose of this paper is to elaborate on how the rocket effect arises due to the drift of material ablated from the pellet and the corresponding asymmetric shielding of the incoming electron heat flux. Building on the plasmoid drift model presented by \citet{vallhagen_drift_2023}, the corresponding shielding length variation across the field lines is evaluated geometrically.
We close the article by providing quantitative predictions for the pellet penetration depth in scenarios representative of a medium-sized tokamak experiment and ITER.


\section{Physical model of the pellet rocket effect}
\label{sec:main-model}

The underlying principle of the pellet rocket effect is that any asymmetric heating of the pellet and the ablation cloud will lead to a higher ablation and pressure on one side of the pellet, yielding a rocket-like propulsion accelerating the pellet towards the less heated side.
In general, this phenomenon involves complex, three-dimensional, non-linear dynamics.

Initial treatments of pellet ablation used the standard NGS model, which made the geometrical assumption of spherically symmetric heating of the neutral cloud. This was soon replaced (see for example \citet{kuteev_hydrogen_1995}) with the more physical assumption for magnetized plasmas, that the incoming electron heat flux is restricted to closely follow the magnetic field lines, so is deposited at the two diametrically opposed regions where the field lines intersect the neutral cloud. The larger thermal ion gyroradius allows heat to be deposited away from these regions \citep{kuteev_hydrogen_1995}, but it will be restricted to the periphery of the cloud \citep{pegourie_2005}, and the neutral cloud pressure is then expected to have a dipole structure, with the maxima along the magnetic field lines. Further improvements beyond the NGS model allowing for the build-up of an electrostatic sheath along the magnetic field line (see for example \citet{kuteev_hydrogen_1995} and \citet{samulyak_2007}), or considering the deformation of the pellet shape \citep{ishizaki_2004}, have indicated that this asymmetry of the lowest-order pressure may be reduced.
The accumulation of physical effects has thus been found to underpin the surprising success of the simple spherically symmetric treatment of ablation dynamics.

Treating the asymmetric pellet ablation dynamics as a linear perturbation around the spherically symmetric dynamics enables us to develop here a semi-analytical model for the pellet rocket effect.
In this section, we consider the momentum transfer at the pellet surface, to connect the perturbed pressure to the force experienced by the pellet.
We also include a discussion of the effect that would be produced by a dipolar correction to the lowest-order neutral cloud pressure.

We take a spherical pellet, of radius $r_\text{p}$, surrounded by ablated neutral gas.
The momentum of ablated particles which leave the pellet surface combines with the gas pressure acting on the pellet surface to generate the force on the pellet. 

\begin{figure}
    \centering
    \includegraphics{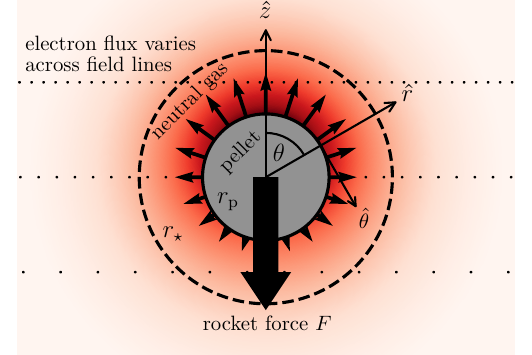}
    \caption{Illustration of how the pellet rocket force arises from both an asymmetry in pressure at the pellet surface (red shading) and an asymmetric ablation (arrows). Additionally, the coordinate system used throughout this paper is indicated. The unit vector $\hat{\vec{z}}$ denotes the axis of asymmetry. $\hat{\vec{r}}$ and $\hat{\vec{\theta}}$ denote the spherical coordinates, in which $\hat{\boldsymbol{\varphi}}$ would point into the paper. The pellet is modelled as a solid sphere of radius $r_\text{p}$.}
    \label{fig:pellet_surface_force}
\end{figure}

Mathematically, the net force on the pellet is calculated by assuming local momentum conservation at each point $\vec{r}$ on the pellet surface $S$ and integrating the stress tensor of the neutral gas as
\begin{equation}
    \vec{F} = - \iint_{S} \left( \rho \vec{v} \vec{v} \cdot \hat{\vec{r}} + p \hat{\vec{r}} \right) \dd{S} \, ,
\end{equation}
with the mass density $\rho(\vec{r})$, the flow velocity $\vec{v}(\vec{r})$ and the pressure $p(\vec{r})$.
The minus sign indicates that this force is exerted on the pellet, while the surface normal unit vector $\hat{\vec{r}}$ points radially outwards.
Let $\hat{\vec{z}}$ point along the axis of asymmetry towards the more strongly heated side, as illustrated in~\cref{fig:pellet_surface_force}.
Note that the illustration assumes no strong dipolar pressure contribution.
The force can then be expressed as
\begin{equation}
    F = - \hat{\vec{z}} \cdot \vec{F} = \iint_{S} \left[ \rho \left(\hat{\vec{z}}\cdot  \vec{v} \vec{v}  \cdot \hat{\vec{r}}\right) + p \left(\hat{\vec{z}}\cdot\hat{\vec{r}}\right) \right] \,\dd{S} \, .
\end{equation}
In spherical coordinates $\{ r, \theta, \varphi \}$, where $\vec{v} = v_r \hat{\vec{r}} + v_\theta \hat{\vec{\theta}} + v_\varphi \hat{\vec{\varphi}}$ and the differential solid angle is $\dd{\Omega} = \sin\theta \dd{\theta}\dd{\varphi}$, the force integral reads
\begin{equation}
    F = r_\text{p}^2 \iint_{S} \left[ \rho v_r (v_r \cos\theta + v_\theta \sin\theta) + p \cos\theta \right] \dd{\Omega} \, .
    \label{Fexpr1}
\end{equation}

Retaining the asymmetries of the quantities appearing in~\cref{Fexpr1} perturbatively allows us to write them as
\begin{align*}
    v_r(\vec{r}) &= v_0(r) + \var{v_r}(r,\theta,\varphi), &
    v_\theta(\vec{r}) &= 0 + \var{v_\theta}(r,\theta,\varphi), \\
    \rho(\vec{r}) &= \rho_0(r) + \var{\rho}(r,\theta,\varphi), &
    p(\vec{r}) &= p_0(r) + \var{p}(r,\theta,\varphi) \, ,
\end{align*}
where subscript $0$ indicates a spherically symmetric component, while perturbations  are denoted by $\delta$. 

Linearizing the force in this perturbation and using $\int_0^\pi \cos\theta \dd{\theta} = 0$ leads to
\begin{equation}
    F = r_\text{p}^2 \iint_S \left[\left( \var{\rho} v_0^2 + 2 \rho_0 v_0 \var{v_r} + \var{p} \right) \cos\theta + \rho_0 v_0 \var{v_\theta} \sin\theta \right] \, \dd{\Omega} \, .
\end{equation}
The last term can be rewritten as a term proportional to $\cos\theta$ through integration by parts, 
which leads to
\begin{equation}
    F = r_\text{p}^2 \iint_S \left[\var{\rho} v_0^2 + 2 \rho_0 v_0  \left( \var{v_r} - {\textstyle \int^\theta_0} \var{v_\theta}(\theta') \dd{\theta'} \right) + \var{p}\right] \cos\theta \dd{\Omega} \, .
    \label{eq:force_before_expansion}
\end{equation}
Consider now that $\cos\theta$ is the $l=1$, $m=0$ mode of the spherical harmonics
\begin{equation}
    Y_l^m(\theta,\varphi) = \sqrt{\frac{(l-m)!}{(l+m)!}} \mathcal{P}_l^m(\cos\theta) e^{im\varphi} \, ,
\end{equation}
with the associated Legendre polynomials $\mathcal{P}_l^m$ and which are orthogonal in the sense
\begin{equation}
    \iint_S Y_l^m (Y_{l'}^{m'})^* \dd{\Omega}=\frac{4 \pi}{2l+1} \delta_{ll'} \delta_{mm'} \, ,
\end{equation}
where  $\delta_{ij}$ denotes the Kronecker delta.
The integral in~\cref{eq:force_before_expansion} can thus be immediately evaluated when expanding the asymmetric perturbations as
\begin{align*}
    \var{\rho}(r,\theta,\varphi) &= \rho_1(r)\cos\theta + \dots \, , & 
    \var{p}(r,\theta,\varphi) &= p_1(r)\cos\theta + \dots \, , \\
    \var{v_r}(r,\theta,\varphi) &= v_{1,r}(r)\cos\theta + \dots \, , &
    \int^\theta_0 \var{v_\theta}(r,\theta',\varphi)\mathrm{d}\theta' &= v_{1,\theta}(r)\cos\theta + \dots \, .
\end{align*}
Higher-order spherical harmonics do not contribute to the net force on the pellet, which can thus be calculated as
\begin{equation}
F = \frac{4 \pi r_\text{p}^2}{3} \left( \rho_1 v_0^2 + 2 \rho_0 v_0 (v_{1,r} - v_{1,\theta}) + p_1 \right)_{r=r_\text{p}} \, .
    \label{eq:rocket_force_full}
\end{equation}
This formula can also be understood physically.
The ablation rate per unit area $g$, i.e. the mass flux through the pellet surface, is
\begin{equation*}
    g(\theta) = \rho \vec{v} \cdot \hat{\vec{r}} \approx \left( \rho_0 v_0 + (\rho_1 v_0 + \rho_0 v_{1,r}) \cos\theta + \dots \right)_{r=r_\text{p}} = g_0 + g_1 \cos\theta + \dots \, .
\end{equation*}
Therefore, the first two terms in~\cref{eq:rocket_force_full} describe the force arising from asymmetric ablation. The third term corresponds to a force from a flow around the pellet surface. The last term $p_1$ describes the gas pressure asymmetry. 

Under the self-regulating shielding assumptions of the NGS model, the flow velocity at the pellet surface is zero \citep{parks_effect_1978}.
Therefore, the pellet rocket force is predominantly caused by the pressure asymmetry at the pellet surface $p_1(r_\text{p})$ and
\begin{equation}
    F \approx \frac{4 \pi r_\text{p}^2}{3} p_1(r_\text{p}) \, .
    \label{eq:rocket_force_short}
\end{equation}
The neglected terms are of the order of the square of the perturbed Mach number at the pellet surface.
A similar expression was assumed in the empirical model developed by \citet{szepesi_radial_2007}.

Pellets will not be perfectly spherical, which we have assumed so far. Effects such as interaction with the wall of the guide tube directing the pellet to the plasma~\citep{ishizaki_2004} or tumbling motion of the pellets~\citep{macaulay_1994} may be expected to smooth the pellet shape. Calculating the expansion dynamics of the ablation cloud from an arbitrarily shaped pellet will be quite difficult, but it is possible to say something definite
about pellets that are almost, but not quite, spherical, using the perturbative approach set out above. Consider a pellet whose boundary in spherical coordinates is given by
\begin{equation}
    r(\theta,\varphi) = r_\text{p} \left(1+\sum_{l,m} \epsilon_{l,m}\mathcal{P}_l^m(\cos\theta) e^{im\varphi} \right)\, ,
    \label{eq:non_spherical_radius}
\end{equation}
where the coefficients in the Fourier sum are small, $\epsilon_{l,m} \ll 1$. If such a pellet is injected into an inhomogeneous plasma, the rocket effect can be calculated approximately by expanding the hydrodynamic equations in two small quantities: the asymmetry of the pellet surface and that of the ablation-cloud heating.
Thanks to the linearity, the effects are additive to lowest order in the expansion.
To this order, it is therefore sufficient to calculate the rocket force on a non-spherical pellet injected into a homogeneous plasma.
The spherical symmetry of the ablation cloud is now broken by the boundary condition at the pellet surface, which is not symmetric. All unknown quantities can be expanded in spherical harmonics, and, thanks to linearity, the coefficients will be proportional to their counterparts in~\cref{eq:non_spherical_radius}. Since the rocket force in the $\hat{\vec{z}}$-direction only depends on the harmonic $(l,m) = (1,0)$, we conclude that $F =c \epsilon_{1,0}$, where the coefficient $c$ can be determined by a simple physical argument. A pellet whose boundary is perturbed by this term only, $r(\theta,\varphi) = r_\text{p} \left(1+\epsilon_{1,0}\cos\theta\right)$, is still spherical, having a boundary that is simply displaced in the z-direction by the constant distance $\epsilon_{1,0}r_\text{p}$. As such, it is clear that the rocket force must vanish, $c = 0$. Thus the departure from perfect spherical symmetry does not affect the rocket force to first order in the smallness of the perturbation.

When we have the situation illustrated in~\cref{fig:pellet_surface_force}, the rocket force will affect the motion of the pellet across the magnetic field lines. This is the case considered in \cref{ssec:results}, where the impact of the rocket force on pellet penetration into a tokamak plasma is analysed. In this case, we note from the last term of~\cref{Fexpr1} that any dipolar structure of the lowest-order pressure such as discussed at the beginning of this section will not affect the rocket force. This pressure structure can introduce asymmetry in the lowest-order ablation flows, but the flows will remain weak (subsonic) near the pellet surface, so no significant contribution would be expected from the first term in~\cref{Fexpr1}.

Another case of interest is when the pellet is heated asymmetrically along the magnetic field lines, so the $\hat{\vec{z}}$ axis is aligned along the field direction and the rocket force drives the pellet along the field lines. This can be the case when multiple pellet fragments exist in close proximity on the same field line, but in this particular situation we also expect a dipolar structure of the lowest-order pressure to develop and to have a significant effect. This is beyond the scope of the semi-analytic model presented here.

In response to the rocket force~\cref{eq:rocket_force_short}, a pellet with mean density $\rho$ will experience an acceleration 
\begin{equation}
 \dot{v}_z = - \frac{p_1}{\rho r_p},
\end{equation}
which will significantly affect the trajectory over the time $\Delta t$ if $|\dot{v}_z| > v_z / \Delta t$. In particular, if the pellet travels the distance $L$ in the $z$-direction, the rocket-force is important if $|\dot{v}_z L / v_z^2| > 1$, i.e. 
\begin{equation}
    p_1 >  \frac{\rho r_p v_z^2}{L}. 
\end{equation}
For typical pellet parameters $v_z \sim 10^3$ m/s, $r_p \sim 10^{-3}$ m, $L = 1$ m, and $\rho \sim 10^2$ kg/m$^3$, a pressure asymmetry as small as 1 atm is thus significant. This value is far smaller than the typical pressure of an ablation cloud, which justifies the linearization undertaken above. 

The major challenge in modelling the pellet rocket effect is to calculate $p_1$ given the pellet and background plasma parameters. An important contribution to the pressure asymmetry is the variation of the plasmoid shielding across the magnetic field, which will be described in the next section.


\section{Shielding asymmetry due to plasmoid drift}
\label{sec:plasmoid-model}

Hot electrons incoming from the background plasma deposit their energy along their path towards the pellet, heating both the neutral cloud and the ionized plasmoid.
Even though the major part of the heating occurs in the neutral gas close to the pellet, the energy deposition in the plasmoid is important in introducing an asymmetry of the heat flux reaching the neutral cloud.
This \emph{plasmoid shielding} of the incoming electron heat flux depends mainly on the integrated density along the electron path through the plasmoid.

In \cref{ssec:plasmoid-geometry} we quantify how the drift of the ionized ablation material leads to a varying shielding length across the magnetic field lines, which in turn affects the pellet ablation dynamics.
In \cref{ssec:plasmoid-shielding} we describe a model for calculating the effective heat flux and  electron energy arriving at the neutral cloud and their asymmetries, which will later be used as boundary conditions for the neutral ablation cloud dynamics.


\subsection{Geometry of the plasmoid boundary}
\label{ssec:plasmoid-geometry}

The flow velocity at the boundary of the neutral ablation cloud rapidly drops 
as the material begins to be ionized~\citep{ishizaki_fluid_2003,pegourie_review_2007,samulyak_2007,bosviel_2021}. Beyond this, at the ionization radius $r_\text{i}$ (not to be confused with the sonic radius $r_\star$ much closer to the pellet), the ablated material can be considered fully ionized and forms the plasmoid ablation cloud.
The dominant dynamics of the plasmoid is its expansion along a flux-tube at the speed of sound 
\begin{equation}
    c_\text{s} = \sqrt{\frac{(\gamma_e \langle Z \rangle + \gamma_i) T_\text{pl}}{\langle m_i \rangle}} \, ,
\end{equation}
where $T_\text{pl}$  is the plasmoid temperature, $\langle m_i \rangle$ and $\langle Z\rangle$ denote the (density weighted) average ion mass and charge number, respectively, and the adiabatic indices of electrons and ions  are $\gamma_e = 1$, $\gamma_i = 3$ \citep{vallhagen_drift_2023}.

\begin{figure}
    \centering
    \includegraphics[width=\textwidth]{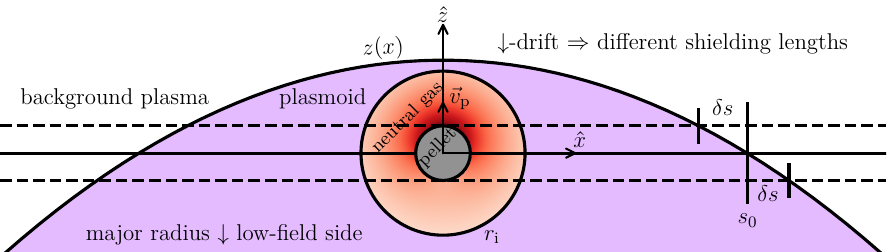}
    \caption{Illustration of the plasmoid shielding asymmetry of the ablation cloud in~\cref{fig:pellet_surface_force}. The drift towards the low-field side (negative $z$ direction) induces a shorter shielding length at the high-field side. Note that the proportions in the figure are not realistic: The pellet is much smaller than the neutral cloud that, in turn, is much smaller than the plasmoid shielding length.}
    \label{fig:plasmoid_shielding}
\end{figure}

Let us consider the plasmoid dynamics in the instantaneous frame of a pellet injected in the opposite direction to the plasmoid drift -- that is, towards the high-field side in tokamaks. The ionized ablated material initially moves at the lab frame pellet speed $v_\text{p}$ towards the low-field side.
The $\vec{E} \times \vec{B}$-drift gradually accelerates material across the field  in the same direction and the plasmoid bends outwards compared to the flux surfaces, as illustrated in~\cref{fig:plasmoid_shielding}.
For a short time following ionization, the plasmoid material undergoes a constant acceleration \citet{vallhagen_drift_2023}
\begin{equation}
    \dot{v}_\text{pl} = \frac{2 (1+ \langle Z \rangle)}{\langle m_i \rangle R_\text{m}} \left( T_\text{pl} - \frac{2n_\text{bg}}{(1+ \langle Z \rangle)n_\text{pl}} T_\text{bg} \right), 
\end{equation}
 where $T_\text{bg}$ and $n_\text{bg}$ are the electron temperature and density of the background plasma, $n_\text{pl}$ is the electron density in the plasmoid, and  $R_\text{m}$ is the local value of the major radius.
The plasmoid electron density can be estimated using particle conservation considerations 
\begin{equation}
    n_\text{pl} = \frac{ \mathcal{G}}{2 \langle m_i \rangle c_\text{s} \pi r_\text{i}^2},
    \label{npl}
\end{equation}
where the outflow from the pellet through a cross-sectional area $\pi r_\text{i}^2$ is determined by the mass ablation rate $\mathcal{G}$. 

Since the particles, once ionized, stop their motion in the positive $\hat{\vec{z}}$-direction nearly instantaneously, the plasmoid boundary $z(x)$, as depicted in~\cref{fig:plasmoid_shielding} with the pellet at the origin, is determined by the trajectory of particles which are ionized at $\{x=0, z=r_\text{i}\}$.
Very approximately, these particles follow the equation of motion
\begin{equation}
    \vec{r}(t) = (\pm c_\text{s} t) \hat{\vec{x}} + \left(r_\text{i} - v_\text{p} t - \frac{1}{2} \dot{v}_\text{pl} t^2 \right) \hat{\vec{z}} \, . 
    \label{rtexpr}
\end{equation}
The shielding length $s(z)$ along a field line at position $z$ is the distance from the plasmoid boundary to the neutral ablation cloud boundary, which lies at ${x = \pm \sqrt{r_\text{i}^2 - z^2}}$.
Therefore, assuming constant acceleration, the shielding length can be estimated by eliminating the time dependence from~\cref{rtexpr}, yielding
\begin{equation}
    s(z) = c_\text{s} \left( - \frac{v_\text{p}}{\dot{v}_\text{pl}} + \sqrt{\left(\frac{v_\text{p}}{\dot{v}_\text{pl}}\right)^2 + \frac{2}{\dot{v}_\text{pl}}( r_\text{i} - z)} \right) - \sqrt{r_\text{i}^2 - z^2} \, .
    \label{eq:shielding_length_full}
\end{equation}
The spherically symmetric part of the ablation is given  by the central shielding length, 
\begin{equation}
    s_0 = s(z=0) = c_\text{s}  \left( - \frac{v_\text{p}}{\dot{v}_\text{pl}} + \sqrt{\left(\frac{v_\text{p}}{\dot{v}_\text{pl}}\right)^2 + \frac{2}{\dot{v}_\text{pl}} r_\text{i}} \right) - r_\text{i} \, . 
    \label{eq:central_shielding_length}
\end{equation}
The asymmetry in the neutral ablation cloud heating is determined by the shielding length variation, $\var{s}$, across the field lines hitting the pellet.
It might appear a reasonable choice to consider the shielding length variation across the entire neutral ablation cloud, however, this would largely overestimate the degree of asymmetry.
The reason is that the vast majority of the heat deposition happens close to the pellet surface, thus the field lines relevant for the asymmetry are the ones spanning the range $z \in [-\var{r}, \var{r}]$, with $r_\text{p} \leq \var{r} \lesssim 1.5 r_\text{p}$. Due to the geometric approximation employed by the NGS model $-\vec{\nabla}\cdot\vec{q} \approx \partial q / \partial r$, which is detailed further in \cref{ssec:asymmetric-ngs-model},
these field lines should correspond to the full angular range $\theta \in [0, \pi]$.
As the shielding length varies nearly linearly across the narrow $z$ range of interest, we may write
\begin{equation}
    \var{s}(z) = \eval{\dv{s}{z}}_{z=0} \!\!\!\! \cdot \var{z} = \eval{\dv{s}{z}}_{z=0}  \!\!\!\! \cdot  \var{r} \cos{\theta} \, , \label{eq:shielding_length_variation}
\end{equation}
with
\begin{equation}
    \eval{\dv{s}{z}}_{z=0} = \frac{- c_\text{s}}{\sqrt{v_\text{p}^2 + 2 \dot{v}_\text{pl} r_\text{i}}} \, . \label{eq:shielding_length_derivative}
\end{equation}


\subsection{Shielding of hot electrons through the plasmoid}
\label{ssec:plasmoid-shielding}

The full dynamics of hot electrons losing energy while traversing a colder plasma involves multiple different mechanisms.
A rigorous description of this heat flux shielding of the plasmoid is ultimately kinetic, and it is outside the scope of this paper.
Here, we approximate the heat flux attenuation by assuming that electrons with a mean free path, $\lambda_\text{mfp}$, shorter than the path length, $d$, they travel through the plasmoid, fully deposit their energy inside the plasmoid. Those electrons satisfying $\lambda_\text{mfp}>d$ are, on the other hand, assumed to be completely unaffected by the plasmoid and retain their thermal kinetic energy from the background plasma.

The path length of an electron trajectory $d$, as it traverses the shielding length $s$, is 
\begin{equation}
    d = \frac{s}{\xi} 
    \, ,
\end{equation}
where $\xi=v_\|/v_e$ is the pitch-angle cosine defined by the electron speed
$v_e$ and its component, $v_\|$, parallel to the magnetic field.  
The mean free path of hot electrons in the plasmoid is determined by their collisions with the cold and dense electron population of the plasmoid 
\begin{equation}
    \lambda_\text{mfp} = \frac{v_e}{\nu_{ee}} = \frac{4 \pi \varepsilon_0^2 m_e^2 v_e^4}{n_\text{pl} e^4 \ln\Lambda} = \left( \frac{v_e}{v_\text{th}} \right)^4 \lambda_T \, , 
    \label{eq:lambda_mfp}
\end{equation} 
with the collision frequency $\nu_{ee}$, 
the cold electron density $n_\text{pl}$ in the plasmoid and the electron mass $m_e$ \citep{helander_collisional_2005}.
For convenience, we define the mean free path $\lambda_T$ at the thermal velocity $v_\text{th} = \sqrt{2T_\text{bg}/m_e}$.
The Coulomb-logarithm  is
\begin{equation}
    \ln\Lambda = \ln\left( \lambda_\text{D} \cdot b_\text{min}^{-1} \right) = \ln \left( \sqrt{\frac{\varepsilon_0 T_\text{pl}}{n_\text{pl} e^2}} \cdot \left( \frac{\langle Z \rangle e^2}{2 \pi \varepsilon_0 m_e v_\text{th}^2} \right)^{-1} \right).
\end{equation}

The condition for an electron to pass through the plasmoid unaffected (i.e. $d<\lambda_\text{mfp}$) can be written in terms of a critical velocity $v_\text{c}$ as
\begin{equation}
    v_e > v_\text{c} \quad \text{with} \quad v_\text{c} = \left( \frac{s}{\xi \lambda_T} \right)^\frac{1}{4} v_\text{th} \, .
\end{equation}
We assume that the background electron distribution is Maxwellian $ f_\text{M}(\vec{v}_e) = ( \sqrt{\pi} v_\text{th} )^{-3} \exp [-(v_e/v_\text{th})^2 ]$.
Then, the heat flux reaching the neutral ablation cloud, which will serve as a boundary condition for the NGS model, can be estimated by integrating $q(\vec{v}_e)$ over the velocities sufficient to pass through the plasmoid,
\begin{gather}
    q_\text{pl} = \underset{\substack{v_e > v_\text{c} \\  \xi \in [0,1]}}{\iiint} 
    \xi v_e \frac{m_e v_e^2}{2} n_\text{bg} f_\text{M}(\vec{v}_e) \dd[3]{v_e} \, .
\end{gather}
This integral can be evaluated in a closed form by introducing ${u = v_e/v_\text{th}} \Rightarrow {u(v_\text{c}) = \left( \xi \alpha \right)^{-\sfrac{1}{4}}}$, with ${\alpha = \lambda_T/s}$. The result is
\begin{equation}
    q_\text{pl}(s) = \underbrace{2 \sqrt{\frac{T_\text{bg}^3}{2 \pi m_e}} n_\text{bg}}_{q_\text{Parks}} 
    \underbrace{\frac{1}{\alpha^2}\left[ e^{\frac{-1}{\sqrt{\alpha}}}\left(-\frac{1}{2}\sqrt{\alpha}+\frac{1}{2}\alpha+\alpha^{3/2}+\alpha^2\right)-\frac{1}{2}\mathrm{Ei}\left(-\frac{1}{\sqrt{\alpha}}\right)\right]}_{f_q(\alpha)} \, , 
    \label{eq:plasmoid_shielding_q}
\end{equation}
with the exponential integral $\mathrm{Ei}(x) = - \int_{-x}^{\infty} \exp(-t)/t \dd{t}$.
Consequently, the heat flux without any plasmoid shielding $q_\text{Parks}$, as assumed by \citet{parks_effect_1978}, is scaled down in our model by the shielding length-dependent dimensionless function, $f_q(\alpha)$.

\begin{figure}
    \centering
    \includegraphics{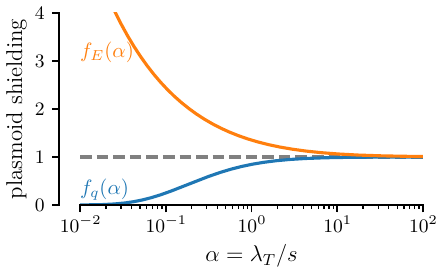}
    \caption{Scaling functions for the heat flux and energy boundary conditions due to plasmoid shielding, inversely dependent on the shielding length $s$.}
    \label{fig:plasmoid_shielding_functions}
\end{figure}

We then define the effective electron energy reaching the neutral ablation cloud as
\begin{equation}
    E_\text{pl} = \frac{q_\text{pl}}{\Gamma_\text{pl}} \, ,
\end{equation}
where the effective particle flux $\Gamma_\text{pl}$ of electrons, defined analogously to $q_\text{pl}$, is 
\begin{equation}
    \Gamma_\text{pl}(s) = \underset{\substack{v_e > v_\text{c} \\  \xi \in [0,1]}}{\iiint} 
    \xi v_e n_\text{bg} f_\text{M}(\vec{v}_e) \dd[3]{v_e} \, .
\end{equation}
The result is again a scaling of the effective energy, $E_\text{Parks}$, assumed by \citet{parks_effect_1978}, by a dimensionless function $f_E(\alpha)$, as
\begin{equation}
    E_\text{pl}(s) = \underbrace{2 T_\text{bg}}_{E_\text{Parks}} 
    \underbrace{\left[
    \frac{ e^{\frac{-1}{\sqrt{\alpha}}}\left(-\frac{1}{2}\sqrt{\alpha}+\frac{1}{2}\alpha+\alpha^{3/2}+\alpha^2\right)-\frac{1}{2}\mathrm{Ei}\left(-\frac{1}{\sqrt{\alpha}}\right)}
    {e^{\frac{-1}{\sqrt{\alpha}}}\left(+\frac{1}{2}\sqrt{\alpha}-\frac{1}{2}\alpha+\alpha^{3/2}+\alpha^2\right) + \frac{1}{2}\mathrm{Ei}\left(-\frac{1}{\sqrt{\alpha}}\right)}
    \right]}_{f_E(\alpha)} \, .
    \label{eq:plasmoid_shielding_E}
\end{equation}
The dimensionless functions $f_q(\alpha)$ and $f_E(\alpha)$ are shown in \cref{fig:plasmoid_shielding_functions}.
For a stronger shielding, i.e. decreasing $\alpha$, the heat flux is reduced, while the effective energy is enhanced, as the shielding filters out a broader range of energies, allowing only more energetic electrons to pass through the plasmoid.

With these shielding length-dependent expressions for the heat flux and effective energy reaching the ablated neutral cloud, we are in the position to provide the boundary conditions for the ablation model. Those for the spherically symmetric component (the basic NGS model) are obtained by evaluating the flux and energy at the central shielding length $s_0$ as
\begin{equation}
\begin{aligned}
    q_\text{bc0} &= q_\text{pl}(s_0) = q_\text{Parks} f_q(\alpha_0),  \\
    E_\text{bc0} &= E_\text{pl}(s_0) = E_\text{Parks} f_E(\alpha_0),
\end{aligned} 
\label{eq:plasmoid_shielding_isotropic}
\end{equation}
with $\alpha_0 = \lambda_T/s_0$ and $s_0$ is given by~\cref{eq:central_shielding_length}.
One subtlety that needs to be addressed is that the plasmoid shielding depends on the ablation rate $\mathcal{G}$ through the plasmoid density $n_\text{pl}$ in \cref{npl}.
However,  $\mathcal{G}$, whilst it has here the spherically symmetric form given by the NGS model (confirmation that our numerical method does recover this is detailed in \citet{guth_pellet_2024-1}), depends itself on the boundary conditions provided by \cref{eq:plasmoid_shielding_isotropic}. Therefore, the plasmoid shielding ($q_\text{bc0}$, $E_\text{bc0}$) has to be calculated in a self-consistent iteration until the ablation rate $\mathcal{G}$ is converged.
Then, the $Y_1^0(\theta,\varphi)=\cos\theta$ components of the flux $q_\text{pl}$ and energy $E_\text{pl}$ provide the boundary conditions for the asymmetric component of the ablation dynamics. 
In the next section, we detail the evaluation of these asymmetric dynamics in the neutral cloud and determine a semi-analytic form for the resulting pressure asymmetry on the pellet surface, in terms of the degree of asymmetry in the boundary conditions.


\section{Asymmetrically heated neutral gas ablation cloud}
\label{ssec:asymmetric-ngs-model}

To determine the pressure asymmetry at the pellet surface, the radial dynamics of the neutral gas cloud must be modelled. We start from the same equations as
the widely used NGS model \citep{parks_effect_1978}, which has been shown to provide reasonable predictions of experimentally observed pellet ablation rates \citep{pegourie_review_2007}.
We outline the details necessary to derive our model here and describe the asymmetric ablation dynamics. For any further details of the spherically symmetric NGS model, we refer the reader to \citep{parks_effect_1978} or \citep{guth_pellet_2024-1}.


\subsection{Asymmetric NGS model}
\label{ssec:asymmetric-NGS}

The neutral gas ablation cloud is considered as a quasi steady-state ideal gas governed by conservation of mass, momentum and energy according to
\begin{align}
    &\frac{\rho}{m} = \frac{p}{T}  &\text{(ideal gas law)} \, , \label{eq:ideal_gas_law} \\
    &\vec{\nabla} \cdot (\rho \vec{v}) = 0 &\text{(mass conservation)} \, , \label{eq:full_mass_conservation} \\
    &\rho (\vec{v} \cdot \vec{\nabla}) \vec{v} = - \vec{\nabla} p &\text{(momentum conservation)} \, , \label{eq:full_momentum_conservation} \\
    &\vec{\nabla} \cdot \left[\left( \frac{\rho v^2}{2} + \frac{\gamma p}{\gamma - 1} \right) \vec{v}\right] = - Q \vec{\nabla} \cdot \vec{q} &\text{(energy conservation)} \, .
    \label{eq:full_energy_conservation}
\end{align}
The fluid quantities describing the gas dynamics at each point $\vec{r}$ are the mass density $\rho$, the pressure $p$, the temperature $T$  and the flow velocity $\vec{v}$.
The gas is taken to consist of ablated molecules of mass $m$ and adiabatic index $\gamma$.
The right hand side of~\cref{eq:full_energy_conservation} represents the heat source and is assumed to be a fraction, $Q \approx \numrange{0.6}{0.7}$, of the loss of energy flux $\vec{q}(\vec{r})$ of the incoming electrons \citep{parks_effect_1978}.
The dominant approximations of the NGS model are in fact the following assumptions made on the form of this heat source. 

First, the energy flux carried by the most energetic electrons is in reality directed along the magnetic field lines, thus passing through the neutral gas cloud in a nearly straight line.
In the NGS model, this is approximated by an equivalent radial path, 
so that $\vec{\nabla}\cdot \vec{q} \approx {\partial q}/{\partial r}$. 
Since most of the heating occurs close to the pellet, this mapping dictates that the asymmetry in $q$ has to be determined by only considering field lines that significantly affect the pellet heating, as discussed in \cref{ssec:plasmoid-geometry}.

Second, the energy distribution of the incident electrons is replaced by a single effective energy $E(\vec{r})$, which we take to be the ratio of the unidirectional energy flux and particle flux facing the pellet cloud.
This enables calculation of the incident electron dynamics through the differential equations
\begin{align}
    &\pdv{E}{r} = 2 \frac{\rho}{m} L(E) &\text{(electron energy loss)} \, , \label{eq:full_electron_energy_loss} \\
    &\pdv{q}{r} = \frac{\rho}{m} q \Lambda(E) &\text{(heat flux attenuation)}  \, , \label{eq:full_effective_heat_flux}
\end{align}
where $L(E)$ and $\Lambda(E)$ are empirical functions describing the energy loss and scattering of electrons passing through a hydrogenic gas \citep{parks_effect_1978}. 

The system of \cref{eq:ideal_gas_law,eq:full_mass_conservation,eq:full_momentum_conservation,eq:full_momentum_conservation,eq:full_energy_conservation,eq:full_electron_energy_loss,eq:full_effective_heat_flux}, together with the required boundary conditions, fully determines the spatial variation of all neutral ablation cloud quantities $y \in \{ \rho, p, T, v_r, v_\theta, q, E \}$.
We separate the spherically symmetric dynamics from the angularly dependent perturbation, as introduced in \cref{sec:main-model}, such that $y(\vec{r}) = y_0(r) + y_1(r) \cos\theta + \dots$ (again with the expansion of $v_\theta$ made on its integral over $\theta$).
\citet{parks_effect_1978} introduced a complete procedure to calculate the spherically symmetric solutions $y_0(r)$, subject here to the symmetric boundary conditions~\cref{eq:plasmoid_shielding_isotropic}, and those are from now on considered to be known functions; see also \citep{guth_pellet_2024-1} for more details.

A set of equations determining the angularly dependent perturbations $y_l(r)$ can then be derived using the linearized versions of \cref{eq:ideal_gas_law,eq:full_mass_conservation,eq:full_momentum_conservation,eq:full_momentum_conservation,eq:full_energy_conservation,eq:full_electron_energy_loss,eq:full_effective_heat_flux}. The perturbative neutral gas dynamics are thus described by 
\begin{align}
    &\rho_l = m \left( \frac{p_l}{T_0} - \frac{p_0}{T_0^2} T_l \right) \, , \label{eq:physical_perturbation_ideal_gas_law} \\
    &\pdv{\rho_0}{r}v_{l,r} + \rho_0 \left[ \frac{1}{r^2} \pdv{r}(r^2 v_{l,r}) - \frac{l(l+1)}{r} v_{l,\theta} \right] + v_0 \pdv{\rho_l}{r} + \frac{1}{r^2}\pdv{r}(r^2 v_0) \rho_l = 0 \, , \label{eq:physical_perturbation_mass_conservation} \\
    &\rho_0 v_0 \pdv{v_{l,r}}{r} + \rho_0 \pdv{v_0}{r} v_{l,r} + v_0 \pdv{v_0}{r} \rho_l = - \pdv{p_l}{r} \, , \label{eq:physical_perturbation_r_momentum_conservation} \\
    &\rho_0 v_0 \pdv{v_{l,\theta}}{r} + \rho_0 \frac{v_0}{r} v_{l,\theta} = - \frac{p_l}{r} \, , \label{eq:physical_perturbation_theta_momentum_conservation} \\
    &\left[ v_{l,r} \pdv{r} + \frac{1}{r^2}\pdv{r}(r^2 v_{l,r}) - \frac{l(l+1)}{r} v_{l,\theta} \right] \left( \frac{1}{2} \rho v_0^2 + \frac{\gamma}{\gamma - 1} p_0 \right) \nonumber \\
    &+ \left[ v_0 \pdv{r} + \frac{1}{r^2}\pdv{r}(r^2 v_0) \right] \left( \frac{1}{2} \rho_l v_0^2 + \rho_0 v_0 v_{l,r} + \frac{\gamma}{\gamma-1}p_l \right) = Q \pdv{q_l}{r} \, , \label{eq:physical_perturbation_energy_conservation}
\end{align}
while the electron dynamics in the neutral gas are described by
\begin{align}
    &\pdv{E_l}{r} = 2 \frac{\rho_l}{m} L(E_0) + 2 \frac{\rho_0}{m} \eval{\pdv{L}{E}}_{E_0} E_l \, , \label{eq:physical_perturbation_electron_energy_loss} \\
    &\pdv{q_l}{r} = \frac{\rho_l}{m} q_0 \Lambda(E_0) + \frac{\rho_0}{m} q_l \Lambda(E_0) + \frac{\rho_0}{m} q_0 \eval{\pdv{\Lambda}{E}}_{E_0} E_l \, . \label{eq:physical_perturbation_effective_heat_flux}
\end{align}
Note that no specific form of the angular dependence was assumed here.
The $\theta$-dependence of the first spherical harmonic ($\propto\cos\theta$) is decoupled from the other modes, due to their orthogonality. Thus, all $y_1(r)$ are described by this self-consistent system of differential equations for $l=1$. 
The only assumption made, apart from the NGS model assumptions, is that the angular dependence is  a small perturbation.

In analogy to the NGS model, the boundary conditions on the pellet surface in the case of hydrogenic pellets are $T_1(r_\text{p}) = 0 = q_1(r_\text{p})$, as the sublimation energy is negligible compared to the thermal energy of the incoming electrons. 
Additionally, the flow velocity vanishes at the pellet surface, as it turns out to be much smaller than the sonic speed.
The incident electrons are treated as coming from $r \rightarrow \infty$, which is motivated by the fact that the vast majority of the heating occurs much closer to the pellet than the ionization radius.

The heating-source boundary conditions are given by $q(r \rightarrow \infty) = q_\text{bc0} + q_\text{bc1} \cos\theta$ and $E(r \rightarrow \infty) = E_\text{bc0} + E_\text{bc1} \cos\theta$.
For convenience, we define the relative asymmetry parameters $q_\text{rel} = q_\text{bc1}/q_\text{bc0}$ and $E_\text{rel} = E_\text{bc1}/E_\text{bc0}$, which are the only input parameters in our ``\emph{asymmetric NGS model}'', in addition to those of the standard spherically symmetric NGS model ($r_\text{p}$, $q_\text{bc0}$, $E_\text{bc0}$). 

The perturbative treatment is justified when $|q_\text{rel}|\ll 1$ and $|E_\text{rel}|\ll 1$.
The signs determine which side of the ablation cloud receives a higher heat flux $q$ or higher effective electron energy $E$.
Note that the sign is not necessarily the same; in fact, the plasmoid drift induced asymmetry described in \cref{sec:plasmoid-model} has typically $E_\text{rel}/q_\text{rel} \approx -1$, as we will see in \cref{ssec:quant-asym}. 

For normalization purposes, we introduce quantities $y_\star = y_0(r_\star)$ taken at the sonic radius $r_\star$, where the continuously accelerated ablated material surpasses the speed of sound of the gas, $c_{s}^{\rm g} = \sqrt{\gamma T_0/m}$. In the rest of the paper, we normalize the spherically symmetric quantities to these values $\widetilde{y}_0 = y_0/y_\star$, while the asymmetric perturbations are normalized as $\widetilde{y}_1 = y_1/(y_\star q_\text{rel}$). 
Details of the numerical procedure to obtain a full solution of the system are provided in Appendix~\ref{numsolApp}.

\begin{figure}
    \centering 
    \begin{subfigure}[b]{0.49\textwidth}
        \centering
        \includegraphics[width=\textwidth]{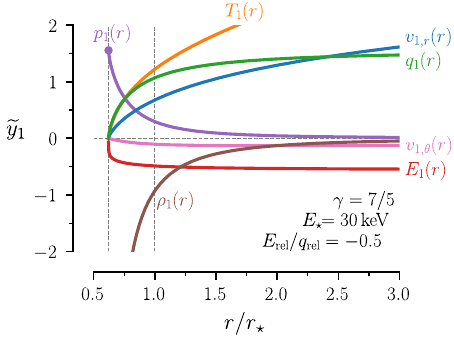}
        \caption{}
        \label{fig:ode1_example_-0.5}
    \end{subfigure}%
    \hfill
    \begin{subfigure}[b]{0.49\textwidth}
        \centering
        \includegraphics[width=\textwidth]{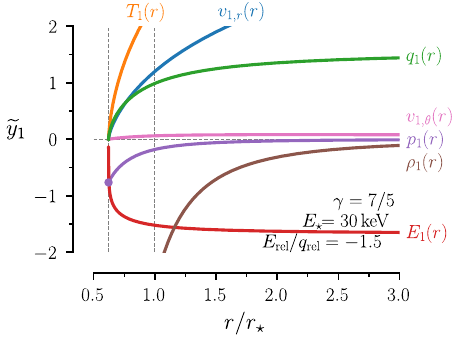}
        \caption{}
        \label{fig:ode1_interesting_-1.5}
    \end{subfigure}
    \caption{Radial dependence of two numerical example solutions to the normalized perturbative ablation dynamics. These solutions correspond to the symmetric NGS model solution shown by \citet{parks_effect_1978} with the parameters $\gamma = 7/5$ and $E_\star(E_\text{bc0}) = \qty{30}{\kilo\eV}$. The heat source asymmetry is characterized here by $E_\text{rel}/q_\text{rel} = -0.5$ in (a) and $E_\text{rel}/q_\text{rel} = -1.5$ in (b). The simulations use $Q=0.65$.}
    \label{fig:example_1st_order_solutions}
\end{figure}

\begin{figure}
    \centering
    \includegraphics[width=\textwidth]{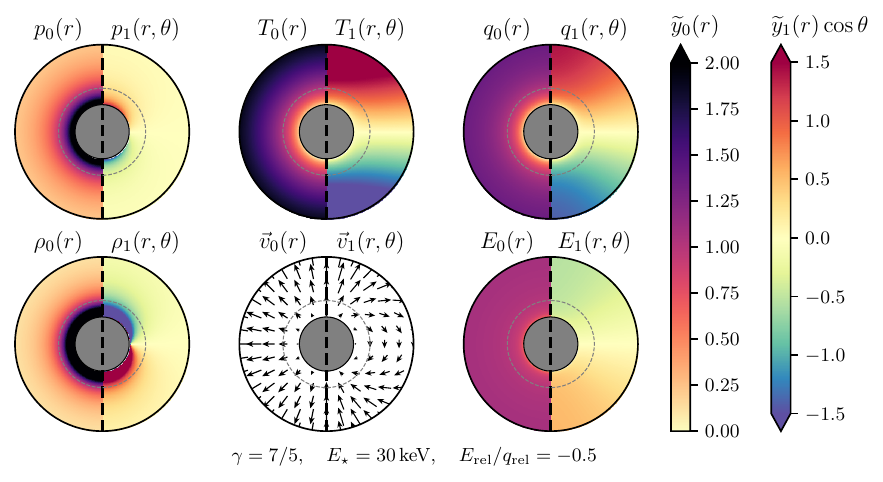}
    \caption{Full spatial dependence of the numerical example solution in \cref{fig:ode1_example_-0.5} for the ablation dynamics. The left sides show the symmetric NGS model solution. The right sides show the perturbative solutions, varying as $\cos\theta$. For illustrative purposes, $v_{1,\theta}$ is scaled up by a factor of 4. The dashed circle marks the sonic radius.}
    \label{fig:full_solution_visuals}
\end{figure}

Two example solutions are shown in \cref{fig:example_1st_order_solutions}.
While this figure shows the radial dependencies of the asymmetric perturbation, it may be easier to interpret the asymmetry through visualization of the full radial and angular dependence.
Therefore, \cref{fig:full_solution_visuals} shows half of the 2D spatial variation of both the symmetric dynamics (on the left) and the asymmetric perturbation (on the right) corresponding to \cref{fig:ode1_example_-0.5}.
The pellet is visualized by the gray circle in the middle, and the dashed line around it indicates the sonic radius.
In reality, the neutral ablation cloud boundary is much further away than visualized.
Scalar quantities are colour-plotted, where darker values correspond to higher absolute values. 
Note that this visualization shows all quantities as their normalized version, while the physical perturbation quantities are much smaller than the spherically symmetric quantities. As discussed in~\cref{sec:main-model}, we do not include any strong dipolar component of the lowest-order pressure $p_0$.

Both the symmetric pressure $p_0$ and density $\rho_0$ are largest close to the pellet surface. On the other hand, the pressure asymmetry $p_1$ exhibits opposite behaviour to the density asymmetry $\rho_1$.
It is evident from the plots of $T$, $q$, $E$ that most of the heating of the neutral gas happens close to the pellet and the asymmetry in temperature $T_1$ follows the asymmetry in heat flux $q_1$.
The flow velocity is presented as a vector field, showing a radial outflow in the symmetric NGS model dynamics, while a flow from the upper side to the lower side is evident in the perturbation.

We find that the dynamics are only weakly sensitive to changes in the adiabatic index $\gamma$ and the energy $E_\text{bc0}$. However, the asymmetry parameter $E_\text{rel}/q_\text{rel}$ can change the dynamics qualitatively.
This can be illustrated by comparing the solutions shown in \cref{fig:ode1_example_-0.5} and \cref{fig:ode1_interesting_-1.5}, corresponding to $E_\text{rel}/q_\text{rel} = -0.5$ and $-1.5$, respectively. In these two cases, the asymmetries in the incoming heat flux and the effective electron energy are of opposite polarity, which is realistic for asymmetries caused by the plasmoid shielding. However, while in the $E_\text{rel}/q_\text{rel} = -0.5$ case the pressure asymmetry is positive, as intuitively expected, it is negative in the $E_\text{rel}/q_\text{rel} = -1.5$ case.
Such a solution corresponds to a reversed rocket force, that is, the pellet accelerates towards the side that receives a higher heat flux.


\subsection{Semi-analytic form for the rocket force}
\label{ssec:rocket-force}

To explore the dependence of $\widetilde{p}_1(r_\text{p})$, and thus the rocket force, on $E_\text{rel}/q_\text{rel}$, we performed a parameter scan, where this quantity is varied along with $\gamma$ and $E_\text{bc0}$.
The dependence on $E_\text{rel}/q_\text{rel}$ is found to be purely linear with a slope depending only weakly on $\gamma$ and $E_\text{bc0}$, as  shown in \citep{GuthPRL}.
Linear regression has produced relative fit errors of $<1\%$, with $|E_\text{rel}/q_\text{rel}|$ ranging from $\num{e-2}$ to $\num{e6}$.
In particular, the linearity also extends into $E_\text{rel}/q_\text{rel} < 0$. 
This linear dependence allows us to construct a simple formula connecting the pressure asymmetry to the degree of asymmetry of the external heating source,
\begin{equation}
    p_1(r_\text{p}) = p_\star a \left(E_\text{rel} -b q_\text{rel} \right) \, .
    \label{eq:p1_at_r_p}
\end{equation}
The fit parameters are found to span the ranges $a \approx 2.0$ to $2.9$ and $b \approx -1.21$ to $-1.17$.
The explicit dependence on $\gamma$ and $E_\text{bc0}$ values is shown in \citep{GuthPRL}.
Corresponding empirical fit functions are provided in the form,
\begin{gather}
    a = a_0 + a_1 \log_{10} E_\text{bc0} \, ,\label{eq:a}\\
    b = b_0 + b_1 (\log_{10} E_\text{bc0})^{b_2} \, ,
    \label{eq:b}
\end{gather}
with coefficients listed in \cref{tab:p1_fit_scaling_laws}.
If there is no asymmetry in the external heat flux, i.e. $q_\text{rel} = 0$, but $E_\text{rel} \neq 0$, the chosen normalization becomes singular.
However, normalization to $E_\text{rel}$ instead can be done similarly and \cref{eq:p1_at_r_p} is found to be valid in this case as well.

\begin{table}
    \centering
    \caption{Coefficients for the scaling laws representing the linear fits of the numerical solutions for the perturbative pressure asymmetry, $p_1(r_\text{p})$.}
    \begin{tabular}{ccccccc}
    \toprule
    $\gamma$ & $a_0$ & $a_1$ & $b_0$ & $b_1$ & $b_2$ \\
    \midrule
    5/3 & 2.1504 & 0.1524 & -1.1622 & -11.3156 & -4.8415 \\
    7/5 & 1.8252 & 0.1066 & -1.1624 & -9.6307 & -4.6843 \\
    9/7 & 1.6999 & 0.0913 & -1.1629 & -11.0180 & -4.7866 \\
    \bottomrule
    \end{tabular}
    \label{tab:p1_fit_scaling_laws}
\end{table}

Finally, our analysis shows that the pellet rocket force indeed depends mainly on $p_1(r_\text{p})$, as anticipated in \cref{sec:main-model}.
Inserting the normalized perturbation quantities into the formula for the pellet rocket force in \cref{eq:rocket_force_full} and using the definition of the sound speed $(\rho_\star v_\star^2 = \gamma p_\star)$ gives
\begin{equation}
    F = \frac{4 \pi r_\text{p}^2}{3} p_\star q_\text{rel} \left( \gamma \widetilde{v}_0^2 \widetilde{\rho}_1 + 2 \gamma \widetilde{\rho}_0 \widetilde{v}_0 (\widetilde{v}_{1,r} - \widetilde{v}_{1,\theta}) + \widetilde{p}_1 \right)_{r=r_\text{p}} \, .
    \label{eq:normalized_force}
\end{equation}
The boundary condition $v_{1,\theta}(r_\text{p}) = 0$ together with zeroth order mass conservation, $r^2 \rho_0 v_0 = \textit{const.}$, results in the corresponding term in the force being zero.
Asymptotic analysis on the other terms is non-trivial, since $v_0(r_\text{p}) \rightarrow 0$ and $v_{1,r}(r_\text{p}) \rightarrow 0$ but $\rho_0(r_\text{p})\rightarrow\infty$ and $\rho_1(r_\text{p})\rightarrow\infty$.
Therefore, this analysis has to be performed numerically, which was done in \citep{guth_pellet_2024-1}, showing that 
the pressure asymmetry $\widetilde{p}_1$ is the dominating term in \cref{eq:normalized_force}, while adding the other terms changes the result only in the third significant figure at most.
 
The expression for the pellet rocket force, with~\cref{eq:p1_at_r_p}, then reads
\begin{equation}
    F = \frac{4 \pi r_\text{p}^2}{3} p_\star \left( a E_\text{rel} - a b q_\text{rel} \right) \, ,
\label{eq:final_pellet_rocket_force}
\end{equation} where $p_\star$ is given by
\begin{gather}
    p_\star = \underbrace{ \frac{\lambda_\star}{\gamma} \left( \frac{(r_\text{p}/r_\star) (\gamma-1)^2}{4 (q_{{\rm bc}0}/q_\star)^2} \right)^\frac{1}{3} }_{f_p(E_{\text{bc}0}, \gamma)} \left[ \frac{m (Q q_{\text{bc}0})^2}{\alpha_\star \, r_\text{p}} \right]^\frac{1}{3} \left(\frac{E_\text{bc0}}{\unit{\eV}} \right)^\frac{1.7}{3} \, ,
    \label{eq:pstar}
\end{gather}
with $\lambda_\star=\rho_\star r_\star \Lambda(E_{\star})/m$,  $\alpha_\star = (E_\text{bc0}/\unit{\eV})^{1.7} \Lambda(E_{\star})$ \citep{parks_effect_1978,guth_pellet_2024-1} and $a$ and $b$ given in \cref{eq:a,eq:b}.
The variables $f_p \approx 0.15$ and $\alpha_\star \approx \qty{1.1e-16}{\m^2}$ vary only weakly with $\gamma$ and $E_\text{bc0}$ and can be considered as constants \citep{GuthPRL}.
The pressure at the sonic radius, $p_\star$, is fully determined by the spherically symmetric heating boundary conditions, $E_\text{bc0}$ and $q_\text{bc0}$, as given by \cref{eq:plasmoid_shielding_isotropic}.
Thus, the rocket force experienced by the pellet is set by the sources of asymmetry in the boundary conditions, $E_\text{rel}$ and $q_\text{rel}$, which is the topic of the next subsection.


\subsection{Quantification of the degree of asymmetry}
\label{ssec:quant-asym}

The degree of asymmetry in the heating of the neutral ablation cloud depends on the shielding length variation given in \cref{eq:shielding_length_variation,eq:shielding_length_derivative}.
However, an additional source of asymmetry is the spatial inhomogeneity of the background plasma parameters, which occurs even in the absence of plasmoid shielding.
Consider the first order variation of \cref{eq:plasmoid_shielding_q},
\begin{align}
    \var{q_\text{pl}}(z) &= \eval{\dv{q_\text{pl}}{z}}_{z=0} \var{z} \\
    &= \left[ \pdv{q_\text{pl}}{T_\text{bg}} \dv{T_\text{bg}}{z} + \pdv{q_\text{pl}}{n_\text{bg}} \dv{n_\text{bg}}{z} + \pdv{q_\text{pl}}{\alpha} \left( \pdv{\alpha}{T_\text{bg}} \dv{T_\text{bg}}{z} + \pdv{\alpha}{s} \dv{s}{z} \right)  \right]_{z=0} \var{z} \\
    &= \left[ \frac{3}{2}\frac{q_\text{pl}}{T_\text{bg}} \dv{T_\text{bg}}{z} + \frac{q_\text{pl}}{n_\text{bg}} \dv{n_\text{bg}}{z} + q_\text{Parks} f'_q \left( \frac{2 \alpha}{T_\text{bg}} \dv{T_\text{bg}}{z} - \frac{\alpha}{s} \dv{s}{z} \right)  \right]_{z=0} \var{z} \, ,
    \label{eq:q_pl_variation}
\end{align}
where $f'_q$ denotes $\partial f_q/\partial \alpha$.
The heat flux asymmetry thus depends on the background temperature and density gradients. We derive $\partial \alpha/\partial T_\text{bg}$ from the definition of $\lambda_T$ in~\cref{eq:lambda_mfp} alone, neglecting any weak temperature dependence of the shielding length or Coulomb logarithm.

As described in \cref{ssec:plasmoid-geometry}, we let $\var{z} = \var{r} \cos\theta$, where $r_\text{p} \leq \delta r \lesssim 1.5 r_\text{p}$, where the upper limit stems from the observation that most of the neutral gas cloud heating is inside the sonic radius $r_\star \approx 1.5 r_\text{p}$ \citep{parks_effect_1978}.
Consequently, the variation $\var{q_\text{pl}}$ corresponds to $q_\text{bc1}\cos\theta$ in our asymmetric NGS model.
The heat flux asymmetry parameter is then
\begin{equation}
    q_\text{rel} = \frac{q_\text{bc1}}{q_\text{bc0}} = \frac{1}{q_\text{pl}(s_0)} \eval{\dv{q_\text{pl}}{z}}_{z=0} \delta r \, .
\end{equation}
Inserting the expressions of \cref{eq:plasmoid_shielding_isotropic,eq:q_pl_variation} and performing an equivalent derivation for the effective energy asymmetry finally gives
\begin{equation}
\begin{aligned}
    q_\text{rel} &= \delta r \left[ \left(\frac{3}{2} \frac{1}{T_\text{bg}} + \frac{f'_q}{f_q}  \frac{2 \alpha}{T_\text{bg}} \right) \dv{T_\text{bg}}{z} + \frac{1}{n_\text{bg}} \dv{n_\text{bg}}{z} - \frac{f'_q}{f_q} \frac{\alpha}{s} \dv{s}{z} \right]_{z=0}, \\
    E_\text{rel} &= \delta r \left[ \left( \frac{1}{T_\text{bg}} + \frac{f'_E}{f_E} \frac{2 \alpha}{T_\text{bg}} \right) \dv{T_\text{bg}}{z} - \frac{f'_E}{f_E} \frac{\alpha}{s} \dv{s}{z} \right]_{z=0}.
\end{aligned}
\label{eq:plasmoid_shielding_asymmetry}
\end{equation}
The plasmoid shielding functions $f_q(\alpha)$ and $f_E(\alpha)$, with $\alpha = \lambda_T/s_0$, are given by \cref{eq:plasmoid_shielding_q,eq:plasmoid_shielding_E}, and visualised in \cref{fig:plasmoid_shielding_functions}. The shielding length $s_0$ and its variation $\eval{\dd s/\dd z}_{z=0}$ are described by \cref{eq:central_shielding_length,eq:shielding_length_derivative}.
Note, that in this model, $q_\text{rel}$ is always positive, however
 $E_\text{rel}$, and thus $E_\text{rel}/q_\text{rel}$, can be negative in the presence of weaker temperature gradients.

 The effect of the background gradient terms alone, neglecting plasmoid shielding, was presented in \citep{GuthPRL}. In this case, $E_\text{rel}/q_\text{rel}$ is always positive, and for parameters characteristic of current experiments, the predicted pellet deceleration is of the experimentally observed order of magnitude, but somewhat smaller. 
 
 When the background gradient terms are negligible compared to the shielding-induced asymmetry terms in \cref{eq:plasmoid_shielding_asymmetry},  $E_\text{rel}/q_\text{rel}$ reduces to $(f_E'/f_E)/(f_q'/f_q)$, which can be calculated from \cref{eq:plasmoid_shielding_q,eq:plasmoid_shielding_E}. In this limit $E_\text{rel}/q_\text{rel}$ is negative and monotonically decreasing with increasing $\alpha$. For $\alpha>1.36$, $E_\text{rel}/q_\text{rel}<-1.17$, which corresponds to $p_1<0$, and thus to the onset of the reverse rocket effect mentioned previously. In particular, in the long mean free path limit $\alpha\gg 1$ the shielding-induced asymmetry alone would cause a reverse rocket effect according to our model.

The effect of the negative $E_\text{rel}/q_\text{rel}$ may however be exaggerated by the monoenergetic electron beam approximation employed inside the neutral cloud. The reason for $E_\text{rel}$ becoming negative in this model is that, in the presence of a reduced shielding, slower electrons are able to penetrate the plasmoid, reducing the average electron energy reaching the neutral cloud. Within the monoenergetic model, a lower pressure inside the neutral cloud is thus sufficient to completely shield the pellet surface from the heat flux. Note that in this case, the energy of the monoenergetic electron population continuously drops inside the neutral cloud. In reality, however, the energy loss rate of more energetic electrons is lower, thus they can reach deeper inside the neutral cloud. Shielding out the heat flux carried by these energetic electrons requires a higher pressure close to the pellet surface. This reduces the impact of the energy asymmetry of the electrons incoming to the neutral cloud, making it more difficult to realize a reverse rocket effect in practice. In this sense, the monoenergetic approximation can be considered as a lower bound on the rocket effect. 

An upper bound on the rocket force can instead be found by setting $E_\text{rel}=0$, while keeping the $q_\text{rel}$ given by \cref{eq:plasmoid_shielding_asymmetry}. This means that the energy distributions are assumed to be equal on both sides of the pellet, i.e. the extra electrons penetrating on the less shielded side are added at a higher energy than they actually have. As a result, our model then predicts that a higher pressure is needed to shield the pellet by neutrals on the side with lower plasmoid shielding than in reality, exaggerating the rocket force in the positive direction. The actual rocket force is expected to be between the values predicted by these limiting cases, but can only be more accurately determined by a kinetic treatment of the neutral cloud dynamics, which is out of scope of the present paper.

In summary, the relative heating asymmetries, $E_\text{rel}$ and $q_\text{rel}$, as given by \cref{eq:plasmoid_shielding_asymmetry}, include both the asymmetries from shielding length variations and from plasma parameter variations, and they can directly be used in \cref{eq:final_pellet_rocket_force} to predict the pellet rocket force.
In addition to the common pellet ablation parameters (pellet radius $r_\text{p}$, background plasma temperature $T_\text{bg}$ and density $n_\text{bg}$), the plasmoid shielding depends on the pellet velocity $v_\text{p}$, the pellet position along the major radius of a tokamak $R_\text{m}$ (representing the magnetic field curvature), the temperature gradient $\mathrm{d}T_\text{bg}/\mathrm{d}z$ and the density gradient $\mathrm{d}n_\text{bg}/\mathrm{d}z$.
Furthermore, values have to be given for the plasmoid temperature $T_\text{pl}$ and the ionization radius $r_\text{i}$, since the plasmoid formation is not fully modelled here.
The provided $E_\text{rel}$ estimate serves as a lower estimate for the rocket force (leading to a potentially reversed rocket force), while setting $E_\text{rel}=0$ is an upper estimate.


\section{Pellet penetration in a medium-sized tokamak and ITER}
\label{ssec:results}

Using our model for the pellet rocket force, we can estimate the pellet penetration depth through prescribed background plasma parameter profiles.  We assume the spherical pellet to be injected from the low-field side, along the mid-plane, with an initial pellet radius $r_0$ and initial speed $v_0$. We calculate the local ablation rate, and so the pellet size reduction, along the trajectory self-consistently, as described in~\cref{ssec:plasmoid-shielding}. We can thus determine the radial trajectory of the pellet with the rocket deceleration computed by our model. 

The influence of the ablated material on the background plasma at the position of the pellet is neglected. This is a reasonable simplification for this injection geometry, since the ablated material is expected to be deposited behind the pellet shard due to the plasmoid drift. To avoid that the deposited material is transported ahead of the pellet, the following analysis excludes situations when a large transport event -- such as a thermal quench -- takes place during the pellet ablation. At the position of the pellet, the main effect on the background plasma is the absorption of thermal energy from the flux tube it resides in. Experience with SPI simulations with DREAM \citep{dreamHoppe} indicate that this is a minor effect for reactor-scale machines even for SPI, owing to the large energy reservoir, but could be a source of moderate inaccuracies in smaller devices.

With the aim of illustrating the impact of the rocket effect in current medium-sized experiments, we first consider a scenario with parameters representative of a medium-sized tokamak (MST) experiment. We use plasma parameters which are realistic, while they do not correspond to any specific experiment. The plasma profiles are parametrized as a modified $\rm tanh$ function, typically used for pedestal profile fits \citep{groebner1998}
\begin{equation}
   y(r)=\frac{y_{\rm max}^{\rm ped}}{2}\left[{\rm tanh}\left(\frac{2 (r_{\rm sym}-r)}{\delta r_{\rm ped}}\right)+1\right]+(y_{\rm max}-y_{\rm max}^{\rm ped})\frac{r_{\rm max}^{\rm core}-r}{r_{\rm max}^{\rm core}}\Theta(r_{\rm max}^{\rm core}-r), 
\end{equation}
for quantity $y\in \{T_{\rm bg},n_{\rm bg}\}$, and minor radius $r\in [0, 0.5]\,\rm m$ (only in this equation, not to be confused with the radial variable of the pellet ablation calculation). Specifically, $r_{\rm sym}=0.475\,\rm m$,  $r_{\rm max}^{\rm core}=0.45\,\rm m$, $\delta r_{\rm ped}=0.05\,\rm m$, as well as $T_{\rm bg,max}^{\rm ped}=0.75\,\rm keV$, $T_{\rm bg,max}=2\,\rm keV$ for the temperature and $n_{\rm bg,max}^{\rm ped}=5\times 10^{19} m^{-3}=n_{\rm bg,max}$ for the density profile. In addition, we assume a plasmoid temperature of $T_{\rm pl}=2\,\rm eV$, and the ionization radius is estimated to be $r_{\rm i}= f_{\rm ri} \, r_{\rm i}^{\rm min}$, with the lower bound on the ionization radius, 
\begin{equation}
    r_\text{i}^\text{min} = \sqrt{\frac{\mathcal{G} (4 \gamma T_\text{pl} + \varepsilon_\text{ion} + \varepsilon_\text{diss})}{m_i \pi q_\text{Parks}}} \, ,
    \label{eq:riminapp}
\end{equation}
calculated according to Eq.~(11) in \citep{parks_radial_2000}, with the ionization energy $\varepsilon_\text{ion} \approx \qty{13.6}{\eV/\text{ion}}$ and the dissociation energy $\varepsilon_\text{diss} \approx \qty{2.2}{\eV/\text{ion}}$. 
This lower bound is determined by requiring that a sufficient heat flux reaches the neutral cloud to be able to ionize the entire neutral outflow. We use $f_{\rm ri}=1$, motivated by the experimental comparison presented in Appendix \ref{trajSzepesi}.
This approximation results in $r_\text{i}$ values  consistent with estimates made by \citet{muller_high_2002} and by \citet{matsuyama_neutral_2022}. In all simulations the quantity $Q$ appearing in \cref{eq:full_energy_conservation} is set to $Q=0.65$.

\begin{figure}
    \centering
    \includegraphics[width=0.49\textwidth]{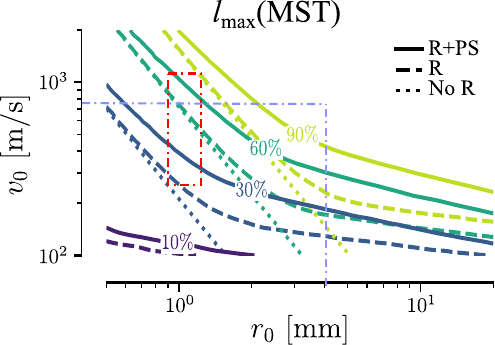}
    \caption{\label{fig:pendepth_asdex} Relative penetration depth as a function of the initial pellet radius $r_0$ and speed $v_0$ in a medium-sized tokamak scenario. Solid lines: with rocket effect, including plasmoid shielding. Dashed: with rocket effect, neglecting the plasmoid shielding. Dotted lines: no rocket effect. Dash-dotted lines indicate representative parameter ranges for fueling and ELM-pacing pellets (red), and SPI fragments (blue). }
\end{figure}

Figure~\ref{fig:pendepth_asdex} shows contours of the relative penetration depth  $l_\text{max}$, as a function of the initial pellet radius $r_0$ and speed $v_0$. This represents the distance reached by the pellet -- before either fully ablating away or turning back due to the rocket force -- normalized to the plasma minor radius.  Lighter tones correspond to higher values of $l_\text{max}$. 
The solid contours (labeled as ``R+PS'') were calculated using our full model for the rocket effect.
To assess the effect of plasmoid shielding induced asymmetries, the dashed contours (``R'') represent the rocket effect from the background plasma variation alone, where the plasmoid shielding is neglected.
The dotted contours (``No R'') are calculated without accounting for the rocket effect, i.e.~considering only the NGS model ablation.

Without the rocket effect (dotted line), the contours of penetration depth are approximately power laws with respect to $r_0$ and $v_0$, i.e. nearly straight lines on this double logarithmic plot.
When including the rocket effect, the penetration depth shows a similar trend for injection velocities $v_0 \gtrsim \qty{500}{\m/\s}$, albeit being somewhat shortened by plasmoid shielding effects (solid line).
The rocket effect induced by plasma parameter variations (dashed line) is negligible in the region of high $v_0$ and low $r_0$, where the dashed and dotted contours coincide.
Towards even higher $v_0$ and smaller $r_0$, even the shielding induced rocket effect becomes less important.

For lower injection velocities and higher pellet sizes, however, where the dashed and solid contours reduce their slope and diverge from the dotted contours, the rocket effect drastically lowers the penetration depth.
Since pellet injection experiments typically operate at higher injection speeds, this explains why the rocket effect -- while detectable -- has not been found to be limiting in current medium-sized experiments \citep{szepesi_2009,kong_interpretative_2024}.
In \citet{GuthPRL}, we evaluated the acceleration resulting from background profile gradients characteristic of those applied here, finding values comparable to the experimental acceleration quoted in \citet{muller_high-_1999} and \citet{muller_high_2002}, but slightly lower.
Generally, as seen in \cref{fig:pendepth_asdex}, including the plasmoid shielding increases the impact of the rocket effect, i.e. shifts the contours towards higher $r_0$ and $v_0$. 
However, in the limit of low $v_0$ and high $r_0$, the dashed contours approach the solid contours and eventually cross them outside the plotted parameter range. 
This crossing is discussed below with regard to the ITER scenarios, where it is clearly visible.

The aim of the analysis shown in \cref{fig:pendepth_asdex} is to explore the parametric dependencies of the rocket effect, including the various asymptotic behaviors, and as such, the parameter ranges plotted are not meant to be representative of any specific experiment. However, as a reference, we indicate parameter ranges relevant for pellets used in normal operation for fueling and ELM pacing in ASDEX-U \citep{plockl2013}, shown with red dash-dotted lines. 
For these pellets the rocket effect is found to be moderate and it is dominated by plasmoid shielding. Note, that fueling pellets are usually injected at the high-field side, whereas the analysis in this section---aiming to provide insights into the penetration depth---assumes low-field side injection. 

We also estimated the parameter range of SPI shards in ASDEX-U (blue dash-dotted), with speed data taken from \citep{Dibon2023}, and upper shard-size estimated from \citep{peherstorfer2022}. Large shards -- which are not numerous, but may contain a significant fraction of the pellet material -- fall in to the region of the parameter space where the total rocket effect is significant and the plasma parameter gradients have a sizable contribution.  We note that while the model does not account for interaction between multiple pellet shards, it provides an indication for the behavior of shards in SPI, particularly those traveling close to the leading edge of the shard plume, where the effect of already deposited pellet material is small.

\begin{figure}
    \centering
        \includegraphics[width=0.49\textwidth]{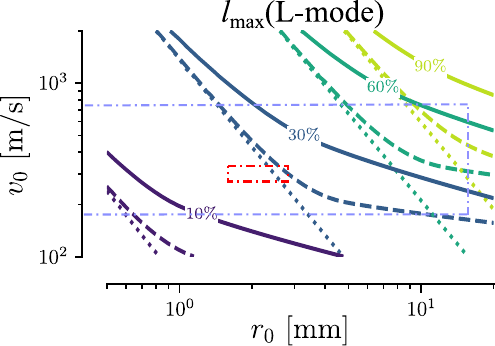}
        \put(-20,125){\small $(a)$}
        \includegraphics[width=0.49\textwidth]{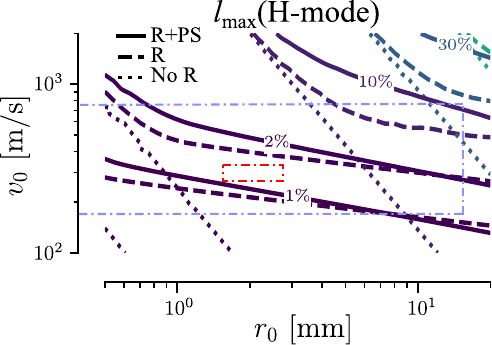}
        \put(-20,125){\small $(b)$}
    \caption{\label{fig:pendepth} Relative penetration depth as a function of the initial pellet radius $r_0$ and speed $v_0$ in two ITER scenarios (a)  L-mode hydrogen plasma (b) H-mode DT plasma. Solid lines: with rocket effect including plasmoid shielding. Dashed: with rocket effect neglecting the plasmoid shielding. Dotted lines: no rocket effect. Dash-dotted lines indicate representative parameter ranges for fueling and ELM-pacing pellets (red), and SPI fragments (blue).}
\end{figure}

We consider two $15\,\rm MA$ ITER scenarios
produced by the CORSICA workflow \citep{kim_investigation_2018}: a low-confinement mode (L-mode) hydrogen plasma ('H26') with a core temperature of $\qty{5.1}{\keV}$ and electron density of $\qty{5.2e19}{m^{-3}}$, and a high-confinement mode (H-mode) DT plasma ('DTHmode24') with a core temperature of $\qty{22.6}{\keV}$ and electron density of $\qty{8.3e19}{\m^{-3}}$ \citep{vallhagen_runaway_2024}. Figure~\ref{fig:pendepth} shows the relative penetration depth as a function of $r_0$ and $v_0$ in the ITER L-mode (a) and H-mode (b) scenarios. Again, we indicate estimated parameter regions for pellets used in normal operation \citep{iterPelletURL} as well as for pellet shards of SPI, with red and blue dash-dotted lines, respectively.  

The behaviour is qualitatively similar to the MST case, except that significantly larger and/or faster pellets are required to achieve the same relative penetration depth, owing to the fact that the ITER plasmas are larger, denser and hotter. In particular, note that, in the H-mode plasma, the pellet cannot penetrate the deep core for the entire parameter range covered.
We observe that the point where the cases with and without the rocket effect diverge takes place at higher values of $v_0$ for higher $r_0$, as well as happening at generally higher values of $v_0$ in the H-mode than in the L-mode plasma. In other words, the relative impact of the rocket effect is higher for most pellet parameters in H-mode than in L-mode. 

A surprising result is the overall limited difference between the cases with and without plasmoid shielding. 
As in the case of a medium-sized tokamak, plasmoid shielding effects dominate for small and fast pellets. Fueling pellets (indicated by the red dashed rectangle) fall into this region of parameter space in L-mode. However, since such small pellets ablate rapidly, they do not have time to decelerate enough before they have completely evaporated to show a significant rocket effect. 

In the region of the parameter space where the rocket effect has a significant impact on the penetration depth -- i.e. towards increasing pellet radii or decreasing velocity -- the background plasma variation is the main source of the rocket effect. In H-mode the fueling pellets fall into this region of parameter space.
Several factors can explain the limited role of plasma shielding here.
First, the relative shielding length asymmetry ($\mathrm{d}s/\mathrm{d}z / s_0$), as given by \cref{eq:central_shielding_length,eq:shielding_length_derivative}, decreases with an increasing ionization radius $r_\text{i}$, which in turn increases with the pellet radius $r_\text{p}$.
It should be pointed out here that the assumed scaling of $r_\mathrm{i}$, as noted before \cref{eq:riminapp}, makes the plasmoid pressure -- and thus the mean free path of the incoming electrons -- independent of $r_\text{p}$. Thus, the $r_0$-scaling of the contribution to the rocket effect from shielding asymmetry is dominated by the scaling of $\mathrm{d}s/\mathrm{d}z / s_0$, rather than the factors $\alpha f_q'/f_q$ and $\alpha f_E'/f_E$ in the last terms in \cref{eq:plasmoid_shielding_asymmetry}.

Second, including plasmoid shielding in our model lowers the overall electron heat flux incident on the neutral cloud and thus lowers the rate of ablation.
With a lower ablation, the pressure asymmetry and in turn the rocket force is lowered.
For high enough $r_0$, this reduction of the ablation even enables the overall impact of the rocket effect to be lower when including plasmoid shielding effects. This is shown by the crossing of the dashed and solid curves in \cref{fig:pendepth}.
The possibility of the reverse rocket effect playing a role here was ruled out by reproducing the penetration depth analysis while enforcing $E_\text{rel} = 0$, yielding nearly identical results.

Moreover, as the rocket force scales rather strongly with the background plasma temperature, most of the deceleration takes place in a relatively short distance towards the end of the pellet trajectory. Thus, even a sizable relative impact of the plasmoid shielding on the rocket force might only be able to affect the pellet speed during a relatively short period, limiting the impact on the penetration depth.

Finally, we note that the choice of how the ionization radius $r_\mathrm{i}$ is treated can have a significant impact on the relative importance of the shielding effect. We have chosen to evolve $r_\mathrm{i}$ along with $r_\mathrm{i}^{\rm min}$, specifically $r_\mathrm{i}=r_\mathrm{i}^{\rm min}$, which yields typical $r_\mathrm{i}$ values of several $\rm cm$ in the ITER scenarios. However, if we reduced the ionization radii by factor of $2$, that is  $r_\mathrm{i}=0.5 r_\mathrm{i}^{\rm min}$, or used a fixed $r_\mathrm{i}=1\,\rm cm $, we would observe significantly reduced penetration depth values for certain parameters when accounting for shielding, especially at high temperatures and small temperature gradients. These latter choices for $r_\mathrm{i}$ would be difficult to physically motivate, thus we do not show corresponding results here, but it nevertheless reflects a notable sensitivity of the shielding effects to $r_\mathrm{i}$. In Appendix~\ref{trajSzepesi} we demonstrate the general agreement between our modelling of the 2D pellet trajectory and the experimental trajectory presented in an example case from~\citet{szepesi_2009}.  


\section{Conclusions}
\label{sec:conclusion}

Cryogenic pellets injected into magnetic fusion devices can be affected by significant acceleration due to asymmetries in the electron heat flux reaching their surface, even when these asymmetries are small. We have rigorously formulated this \emph{rocket effect} problem, where the asymmetry of the electron heat flux -- along with all the asymmetries it generates in the relevant fluid quantities of the neutral cloud ablated from the pellet -- are treated as perturbations around the spherically symmetric neutral gas shielding (NGS) ablation model. We note that strong interactions between pellet fragments affecting their motion along a given magnetic field line is beyond the scope of the current model.

The force acting on the pellet is found to be dominated by the pressure asymmetry in the neutral cloud, while contributions from asymmetric ablation and mass flows in the neutral cloud are negligible. Based on numerical computation of the pressure asymmetry over wide ranges of the incoming electron energy and relevant asymmetry parameters, we provide a simple fitted expression for the rocket force. This expression depends on the relative asymmetries of the characteristic electron energy $E_{\rm rel}$ and the electron heat flux $q_{\rm rel}$, besides the usual inputs of the NGS model.

In order to make the model complete, we derive analytical formulae for $E_{\rm rel}$ and $q_{\rm rel}$, accounting for contributions from radial profile variations of the background plasma and asymmetric plasmoid shielding caused by plasmoid drifts. A peculiar, counter-intuitive prediction of the model is a \emph{reverse rocket effect} for $E_{\rm rel}/q_{\rm rel}<-1.17$, where the rocket force accelerates the pellet in the direction where the heat flux reaching the neutral cloud is higher. We then argue that this corresponds to a lower bound on the rocket effect that is difficult to reach in practice, while an upper bound is given by the choice $E_{\rm rel}=0$.

Finally, we calculate the penetration depth of pellets for plasma profiles representative of a medium-sized tokamak and high plasma current ITER discharges. The rocket effect is shown to have a significant impact on penetration depth for large initial pellet sizes $r_0$ and low initial pellet speeds $v_0$. For hotter plasmas, especially for the H-mode ITER scenario considered, a larger part of the $r_0$-$v_0$ space is affected by the rocket force, which indicates the increased importance of the phenomenon in reactor scale devices. The relative importance of the shielding induced asymmetry is found to depend on the choice of the ionization radius of the pellet cloud. However, for a realistic choice of this parameter it appears that in the regions of the $r_0$-$v_0$ space where the rocket effect has a major impact on penetration depth, it is dominated by the contribution from the asymmetry due to radial profile variations. On the other hand, for low $r_0$ and high $v_0$, the rocket effect is only weakly affecting the penetration depth, but then it is dominated by the asymmetry induced by the plasmoid shielding.   

The model presented here is valid only for hydrogenic pellets, and it relies on  approximations, such as a mono-energetic treatment of electrons, thus it should only be considered as an estimate of the rocket effect. Relaxing some of these approximations presents interesting avenues for future extensions, for example, incorporating kinetic effects into the treatment of energy deposition in the neutral cloud by building on the approach of \citet{fontanillabreizman_2019}. We have demonstrated favourable comparisons between the model and existing experimental results, but a more extensive comparison against high fidelity simulations, such as those of \citet{kong_interpretative_2024}, is beyond the current scope. Nevertheless, even in its current form, the model is useful, as it is sufficiently simple to allow the exploration of large parameter spaces to develop the understanding of the processes at work, and it is suitable for implementation in complex modelling frameworks, such as the disruption modelling tool DREAM \citep{dreamHoppe}. This would allow the plasma response to the pellet ablation to be accounted for in the latter framework, thus high field side injection, as well as multiple or shattered pellet injection scenarios, could be considered in a self-consistent manner.     

\emph{Acknowledgements} --
The authors are grateful to E.~Nardon and P.~Aleynikov for fruitful discussions.  The work was supported by the Swedish Research Council (Dnr.~2022-02862 and 2021-03943), and by the Knut and
Alice Wallenberg foundation (Dnr. 2022.0087 and 2023-0249), and part-funded by the EPSRC Energy Programme [grant number EP/W006839/1]. The work has been partly carried out within the framework of the EUROfusion Consortium, funded by the European Union via the Euratom Research and Training Programme (Grant Agreement No 101052200 — EUROfusion). Views and opinions expressed are however those of the authors only and do not necessarily reflect those of the European Union or the European Commission. Neither the European Union nor the European Commission can be held responsible for them. 


\appendix
\section{Numerical solution of the asymmetric NGS model}
\label{numsolApp}

The normalization introduced, $\widetilde{y}_0 = y_0/y_\star$ and $\widetilde{y}_1 = y_1/(y_\star q_\text{rel}$), is convenient as it reduces the free parameters of the system to only $\gamma$, $E_\text{bc0}$ and $E_\text{rel}/q_\text{rel}$.
The physical solution is then inferred from the normalized numerical solution by additionally providing $r_\text{p}$, $q_\text{bc0}$ and $q_\text{rel}$.

With this normalization, the system of equations~\cref{eq:physical_perturbation_ideal_gas_law,eq:physical_perturbation_mass_conservation,eq:physical_perturbation_r_momentum_conservation,eq:physical_perturbation_theta_momentum_conservation,eq:physical_perturbation_energy_conservation,eq:physical_perturbation_electron_energy_loss,eq:physical_perturbation_effective_heat_flux}
barely changes, with variables acquiring a tilde and the following factors entering the right side of the equations: 
\begin{equation*} 
    {\arraycolsep=1.4pt\def\arraystretch{1.3}
    \begin{array}{cll}
        \frac{1}{m} && \text{in \cref{eq:physical_perturbation_ideal_gas_law},} \\
        \frac{1}{\gamma} &\quad\quad& \text{in~\cref{eq:physical_perturbation_r_momentum_conservation,eq:physical_perturbation_theta_momentum_conservation},} \\
        \frac{1}{\gamma-1} \frac{2}{\lambda_\star Q} && \text{in~\cref{eq:physical_perturbation_energy_conservation} and} \\
        m \lambda_\star &&\text{in~\cref{eq:physical_perturbation_electron_energy_loss,eq:physical_perturbation_effective_heat_flux},}
    \end{array}
    }
\end{equation*}
with the dimensionless NGS model quantity $\lambda_\star = \rho_\star r_\star \Lambda_\star/m$.
On the left side of \cref{eq:physical_perturbation_energy_conservation}, $p_0$ and $p_l$ get an additional factor $1/\gamma$.
The normalized heating boundary conditions are $\widetilde{q}_1(\infty) = \widetilde{q}_0(r\rightarrow\infty)$ and $\widetilde{E}_1(\infty) = \widetilde{E}_0 E_\text{rel}/q_\text{rel}$.

It is convenient for the numerical solution to write the normalized system of differential equations in a matrix form, $\partial \vec{\widetilde{y}}_1/\partial \widetilde{r} = C \vec{\widetilde{y}}_1$, using the computer algebra system SymPy\footnote{\url{https://www.sympy.org/en/index.html}}, where $\vec{\widetilde{y}}_1=(\widetilde{p}_1, \widetilde{T}_1, \widetilde{v}_{1,r}, \widetilde{v}_{1,\theta}, \widetilde{q}_1, \widetilde{E}_1)^T$  and $C$ is a coefficient matrix determined by $\widetilde{y}_0$. When doing this, an apparent singularity appears in the first three rows of $C$, of the form $1/(\widetilde{T}_0 - \widetilde{v}_0^2)$, where $\widetilde{T}_0(\widetilde{r}=1) = \widetilde{v}_0^2(\widetilde{r}=1) = 1$. 
Similar to \citet{parks_effect_1978}, we require that $\left. \partial \widetilde{y}_1/\partial \widetilde{r}\right|_{\widetilde{r}=1}$ is finite, allowing us to apply l'H\^{o}pital's rule to obtain analytical expressions for $\left.C\right|_{\widetilde{r}=1}$. This requirement also eliminates one unknown normalized quantity at $\widetilde{r}=1$, according to
\begin{equation} 
    \widetilde{v}_{1,\theta}(\widetilde{r}=1) = \left[ \left(1- \frac{\chi_\star}{2}\right) \widetilde{v}_{1,r} + \left(1 + \frac{\chi_\star}{4}\right)\widetilde{T}_1 - \widetilde{q}_1 - \pdv{\widetilde{\Lambda}}{E} \widetilde{E}_1 \right]_{\widetilde{r}=1} \, ,
    \label{eq:perturbation_sonic_relation}
\end{equation}
where $\chi_\star = \left. \partial \widetilde{v}_0^2 / \partial \widetilde{r}\right|_{\widetilde{r}=1}$ is given by \citet{parks_effect_1978}. Further details of the above procedure are provided in appendix A.2 of \cite{guth_pellet_2024-1}.

Now that we have an expression for $C$ at every point $\widetilde{r}$ based on $\widetilde{y}_0(\widetilde{r})$, the solution $\widetilde{y}_1(\widetilde{r})$ can be calculated numerically as an initial value problem starting from $\widetilde{r}=1$.
Five of the six $\widetilde{y}_1(\widetilde{r}=1)$ quantities are treated as parameters, which are varied in an optimization scheme until the boundary conditions at $\widetilde{r}_\text{p}$ and $\widetilde{r}\rightarrow\infty$ are satisfied.


\appendix
\addtocounter{section}{1}
\section{Dependence of pellet trajectory on ionization radius}
\label{trajSzepesi}

In~\citep{szepesi_2009}, experimental pellet trajectories in ASDEX-Upgrade were compared to those predicted by two previous models. One in which the pressure asymmetry was determined by experimental comparison, and one in which it results from only plasmoid shielding. They found a range of agreement depending on the plasma scenario considered.

\begin{figure}
    \centering 
        \includegraphics[width=0.49\textwidth]{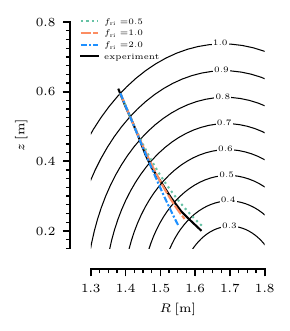}
        \put(-160,30){\large a)}
        \includegraphics[width=0.49\textwidth]{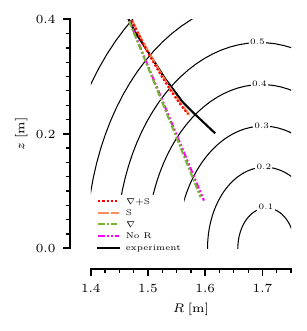}
        \put(-160,30){\large b)}
    \caption{\label{fig:peltraj} Computed pellet trajectories. a) Including the effects of background plasma gradients and plasmoid shielding, for three different values of $f_\mathrm{ri}$: $0.5$ (dotted curve), $1$ (dashed), and $2$ (dash-dotted). b)  Including both gradients and shielding effects (dotted); only shielding (dashed); only gradients (dash-dotted); and without rocket effect (dash-double-dotted). The equilibrium and plasma profiles are based on~\citet{szepesi_2009}, figure 3c. In both panels, the approximate experimental pellet trajectory is indicated by the black curve.
    }
    
\end{figure}

We have constructed a Miller-type equilibrium~\citep{miller_1998} to represent that given in their figure 3c. Using the parametrization 
\begin{equation*} 
    R = R_0 + \Delta(r) + r \cos (\theta + \delta(r) \sin \theta),  \quad\quad
    z = r\kappa(r) \sin \theta,
 \label{flux-surf-param}
\end{equation*}
we could approximate the magnetic geometry choosing $\Delta(r)=-0.03\, r$, $\kappa(r)=1.4+0.075 \, r$, and $\delta=0$. Their normalized flux surface labels satisfy $\psi(r) \approx 2r$. The resulting equilibrium surfaces are shown in~\cref{fig:peltraj}, labeled by $\psi$.

The pellet velocity is taken to be $600\,\rm m/s$ and the pellet radius follows from $r_p = \left(3 m_\text{pel}n_{D}/4\pi N_A\rho_\mathrm{pel}\right)^{1/3}$ as $r_p \approx 1.01 \ \mathrm{mm}$, with $m_\text{pel} \approx 2.62 \times 10^{20}$ in terms of the number of deuterium atoms, $n_{D} \approx 2.01 \,\rm g/mol$ the molar mass of D,  the pellet density $\rho_\mathrm{pel} = 204\,\rm kg/m^3$~\citep{senichenkov_pellet_2007} and $N_A$ Avogadro's constant. We note, that \cref{eq:shielding_length_full,eq:plasmoid_shielding_asymmetry} assume injection along the outboard mid-plane; to be able to use them for the injection geometry considered here, trivial geometrical generalizations were necessary.   

In~\cref{fig:peltraj}a, the computed pellet trajectory, accounting for the background plasma gradients and plasmoid shielding, is shown for three values ($0.5$ dotted curve, $1$ dashed, and $2$ dash-dotted) of the prefactor $f_{\rm ri}$ used to estimate the ionization radius  $r_\mathrm{i} = f_{\rm ri} r_\mathrm{i}^{\rm min}$. The approximate experimental trajectory from~\citet{szepesi_2009} is overlaid in black, and a reasonable agreement can be seen when the ionization radius is close to the minimum, $r_\mathrm{i}^{\rm min}$. This motivates our use of $f_{\rm ri}=1$ in the simulations presented in the rest of the paper. 

In this scenario the rocket force is dominated by the shielding asymmetry, as illustrated in~\cref{fig:peltraj}b. The difference between the cases where we do not account for the rocket force at all (dash-double-dotted curve), and where only the gradient contribution is retained  (dash-dotted curve), is negligible. Similarly, when accounting for the shielding effect, whether we also keep the gradient contribution  (dotted) or not (dashed), makes no visible difference between the trajectories. This observation is consistent with being in the parameter region with small pellet size and high pellet velocity -- corresponding to the upper left quadrant in our medium sized tokamak example,~\cref{fig:pendepth_asdex}.     


The pressure asymmetry factor giving the rocket force in~\citet{szepesi_2009}, their equation (3), can be identified from our~\cref{eq:final_pellet_rocket_force} as $\epsilon \equiv (4/3)\left( a E_\text{rel} - a b q_\text{rel} \right)/3.5$. The factor 3.5, as determined in~\citet{parks_effect_1978}, relates the pressure at the pellet surface to $p_\star$. Along the trajectories shown, the pressure asymmetry initially rises quickly, and falls rapidly near the end of the trajectory. In particular, for the $f_{\rm ri}=1$ case it is fairly steady at around 5\% over most of the trajectory. This compares favourably to the 5-7\% quoted by~\citet{szepesi_2009} to recover the experimental trajectory.


\bibliographystyle{jpp}

\bibliography{references_from_Zotero.bib,references_manually_added.bib}

\end{document}